\documentclass[prb,aps,showpacs,amsmath,amssymb,twocolumn]{revtex4}
\usepackage{graphicx}
\usepackage{epsf}
\begin{document}
\title{Theory of pinning in a Superconducting Thin Film Pierced
by a Ferromagnetic Columnar Defect}
\author{M. Amin Kayali}
\email{amin@hnl.bcm.tmc.edu}
\affiliation{Human Neuroimaging Laboratory, Baylor College of Medicine,
One Baylor Plaza, Houston, TX 77030, USA.}
\begin{abstract}
This is an analytical study of pinning and spontaneous vortex phase
is a system consisting of a superconducting thin film pierced by a
long ferromagnetic columnar defect of finite radius $R$. The
magnetic fields, screening currents, energy and pinning forces for
this system are calculated. The interaction between the magnetic
field of vortices and the magnetization outside the plane of the
film and its close proximity enhances vortex pinning significantly.
Spontaneous vortex phase appears when the magnetization of the
columnar defect is increased above a critical value. Transitions
between phases characterized by different number of flux quanta are
also studied. These results are generalized to the case when the
superconductor is pierced by an array of columnar defects.
\end{abstract}
\maketitle
\section{Introduction}
Optimizing pinning in superconductors (SC) is a problem that is
interesting both experimentally and theoretically since pinning is
the main viable mechanism for superconductivity in the presence of
external magnetic field. Many tools and mechanisms for pinning has
already been studied, in particular the use of crystal defects
such as holes, non-magnetic impurities and both linear and screw
dislocations. Pinning using structural defects suffers from many
drawbacks, the most important is that the pins are randomly
distributed which results in a low critical current. In recent
years, it was claimed that pinning could be optimized if we employ
ferromagnetic (FM) nano textures to pin superconducting vortices (SV).
\cite{schull}-\cite{elliptic}

The main obstacles of using ferromagnetic textures to pin
superconducting vortices are the proximity effects which destroy
superconductivity. Lyuksyutov and Pokrovsky noticed
that proximity effects could be suppressed if a thin layer of
insulator oxide is sandwiched between the FM and the SC.\cite{PL1} Due to
this separation between the FM and SC, the interaction between
them is mediated via their magnetic fields. If the magnetization
of the FM structure exceeds a threshold value $M_c$, the
interaction between the FM and SC makes the spontaneous creation
of superconducting vortices energy favorable. The interaction
between the FM and SC increases as the temperature reaches the
superconducting transition temperature $T_s$ from below.

Marmorkos {\it{et.al.}}\cite{peeters} considered
a system consisting of a thin ferromagnetic dot embedded in a
superconducting thin film. Both the superconductor and the ferromagnet are
assumed to lie in the $xy$-plane and the dot is magnetized along $z$-axis.
By numerically solving the nonlinear Ginzburg-Landau equation, they were
able to show that a vortex appears in the superconductor when the
magnetization of the dot exceeds a critical value. They also showed
that increasing the dot's magnetization leads to a giant vortex state
with multiple flux quanta.

In Ref.\onlinecite{ours}, Erdin {\it{et.al.}}
used London's theory of superconductors to solve the general problem
of the interaction between vortices in superconducting thin film with
a generic two-dimensional ferromagnetic structure. They applied
their results to the cases when the FM structure is a circular dot whose
magnetization is either parallel or perpendicular to the plane of the
superconductor. They calculated the threshold value for the dot's
magnetization at which vortices are spontaneously created in the
superconductor. They showed that by increasing the magnetization of the
dot a series of phase transitions between phases with different number
of vortices take place.

\begin{figure}[t]
  \centering
  \includegraphics[angle=0,width=3.0in,totalheight=1.5in]{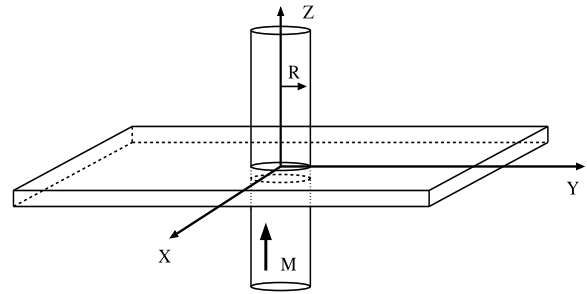}
  \caption{A superconducting thin film pierced by a ferromagnetic nano rod
    of radius $R$, length $2L$ and magnetization $M$.}
  \label{fig1}
\end{figure}

All these studies focused on cases in which the ferromagnet is either
a point-like dipole or an infinitely thin two-dimensional texture whose
plane is parallel to the plane of the superconductor. The problem of
interaction between vortices in superconducting thin film with
a ferromagnetic columnar defect (FCD) has not been
studied yet. This case is both theoretically and experimentally
interesting for the following reason. The magnetic field of the
vortex in a superconducting thin film is not confined to the
plane of the film but it also exist outside of it.
If the ferromagnet extends in space outside the film
and its close proximity then the vortex will be bound more strongly
to the ferromagnet due to the interaction of its magnetic field with
the magnetization of the ferromagnet. This enhances vortex pinning and
consequently the critical current of the superconductor.
It is also important to note that ferromagnetic columnar defects
makes the superconductor has a multiply connected topology which
is another source of pinning. Therefore, it would be interesting
both theoretically and experimentally to study the statical and
dynamical properties of a superconducting thin film pierced by
ferromagnetic columnar defects.

In this article, I propose to study the properties of spontaneous
vortex phase and pinning in a superconducting thin film pierced
by ferromagnetic columnar defects. This article is organized
such as in the first section, the magnetic potential and field
distribution produced by a single FCD penetrating a thin
superconducting film are calculated. In sections
two, I calculate the total energy of a system of a superconducting
vortex coupled to the magnetic defect. I generalize these results
to the case of a superconducting thin film pierced by a square array of
magnetic columnar defects in section four then I summarize the
results of this work in section five.

\section{An SC film pierced by a ferromagnetic columnar defect}
I consider a superconducting thin film in the $xy$-plane pierced by
a finite ferromagnetic nano rod whose radius is $R$ and length is
$L$ as shown in Fig.~\ref{fig1}. I also assume that the FCD is
uniformly magnetized along its symmetry axis then its magnetization
distribution is written as

\begin{eqnarray}
{\bf{M}}(\rho,\varphi,z)=M \Theta(R -\rho)\Theta(\frac{L}{2} -|z|)
\hat{z} \label{Mag}
\end{eqnarray}
\noindent where $\hat{z}$ is the unit vector along the $z$-axis. The
magnetic field produced by the FCD penetrates the superconductor and
changes the distribution of the screening current. In the London
approximation the FCD-SC system is described by London-Maxwell
equation
\begin{eqnarray}
{\bf{\nabla}}\times {\bf{\nabla}} \times {\bf{A}}= \frac{4\pi}{c}
{\bf{J}}
\label{LM}
\end{eqnarray}
The total current ${\bf{J}}={\bf{J}}^s +{\bf{J}}^m$ is the sum of
the supercurrent ${\bf{J}}^s$ and the magnetization current
${\bf{J}}^m=c {\bf{\nabla}}\times {\bf{M}}$. The supercurrent can be
written in terms of the gradient of the superconducting order
parameter $\chi$ and the total magnetic vector potential
${\bf{A}}={\bf{A}}^s +{\bf{A}}^m$ for the FCD-SC system
${\bf{J}}^s=\frac{n_s \hbar e}{2m_e}\left({\bf{\nabla}}\chi
-\frac{2\pi}{\phi_0} {\bf{A}}\right)$. Here $n_s$ is the
Superconducting electron density, $\hbar$ is Planck's constant while
$m_e$ and $e$ are the electron's mass and charge respectively. The
superposition principle allows us to solve (\ref{LM}) for
${\bf{A}}^m$ and ${\bf{A}}^s$ independently. If the Coulomb gauge
${\bf{\nabla}}\cdot {\bf{A}}^m=0$ is imposed then ${\bf{A}}^m$
satisfies the following equation
\begin{eqnarray}
-\nabla^2 {\bf{A}}^m=-\frac{1}{\lambda} {\bf{A}}^m \delta(z)
+4\pi {\bf{\nabla}}\times{\bf{M}}
\label{GLM}
\end{eqnarray}
\noindent where $[\lambda=\lambda_L^2/d_s]$ with
$[\lambda_L=\sqrt{m_e c^2/4\pi n_s e^2}]$ is London's penetration
depth. Here $m_e$ and $e$ are the electron's mass and its charge
while $n_s$ is the superconducting electrons density and $d_s$ is
the thickness of the SC film. The general solution of (\ref{GLM}) is
of the form
\begin{eqnarray}
{\bf{A}}^m=\frac{1}{(2\pi)^3}\int \tilde{{\bf{A}}}_{\bf{k}}^m
e^{-\imath ({\bf{q}}\cdot {\bf{\rho}}+k_z z)} d^2q dk_z
\end{eqnarray}
where $k_z$ and ${\bf{q}}$ are the components of the wave vector along the
$z$-axis and in the $xy$-plane. Here $\tilde{{\bf{A}}}_{\bf{k}}^m$ is
the vector potential in momentum space and is given by
\begin{eqnarray}
\tilde{{\bf{A}}}_K^m =\frac{16\pi^2 \imath M R J_1(qR)}{k_z^2 +q^2} \left[
\frac{2\sinh(\frac{qL}{2})e^{-\frac{qL}{2}}}{q(1+2\lambda q)}
-\frac{\sin(\frac{k_z L}{2})}{k_z}\right]\hat{\varphi}_q
\label{FRVP}
\end{eqnarray}
where $J_n (x)$ is the $n$-th rank Bessel's function.
The  magnetic field produced by the magnetic nano-rod in the presence of
the SC film can be calculated using ${\bf{B}}={\bf{\nabla}}\times{\bf{A}}^m$.
The components of the rod's magnetic field are
\begin{widetext}
\begin{eqnarray}
B_z&=&\frac{8\pi M R}{\lambda}\int_0^\infty J_1(qR)
J_0(q\rho)\left[
\frac{\pi}{2q^2}W(q,z,L)-\frac{\sinh(\frac{qL}{2})
e^{-q(|z|+\frac{L}{2})}}{1+2q}\right]dq \\
B_\rho&=&\frac{8\pi M R}{\lambda}\int_0^\infty J_1(qR) J_1(q\rho)
\left[\frac{\pi}{2q^2}W(q,z,L)-\frac{\sinh(\frac{qL}{2})
e^{-q(|z|+\frac{L}{2})}}{1+2q}\right]dq
\end{eqnarray}
where $W(q,z,L)$ is
\begin{eqnarray}
W(q,z,L)&=& sign(L-2z)\left[1-\cosh(\frac{q(L-2z)}{2})
+sign(L-2z) \sinh(\frac{q(L-2z)}{2})\right]\nonumber\\
&+&sign(L+2z)\left[1-\cosh(\frac{q(L+2z)}{2})
+\sinh(\frac{q|L+2z|}{2})\right]
\label{integ}
\end{eqnarray}
\end{widetext}
However, we are interested in the value of the field at the plane of
the superconductor. The $z$ component of the magnetic field of the
FCD evaluated at the SC plane is
\begin{eqnarray}
B_z^m (\rho)=\frac{8\pi M R}{\lambda}\int_0^\infty \frac{q
J_1(qR)J_0(q\rho)}{1+2q} (1-e^{-\frac{qL}{2}}) dq
\end{eqnarray}
Similarly the solution for ${\bf{A}}^s$ is found. A general argument
made in \cite{ours} shows that the term proportional to
${\bf{\nabla}}\chi$ ascribes for vortices. So, in the presence of a
vortex with vorticity $\nu$ and center at ${\bf{\rho}}_0=0$, the
solution for ${\bf{A}}^s$ in the Coulomb gauge is

\begin{eqnarray}
{\bf{A}}^s ({\bf{\rho}},z)=\frac{\nu \phi_0}{2\pi} \hat{z}\times
\hat{\rho} \int_0^\infty
\frac{J_1(q|{\bf{\rho}}|)e^{-q|z|}}{1+2\lambda q} dq
\end{eqnarray}
where $\hat{\rho}$ is the unit vector along ${\bf{\rho}}$. The
$z$-component of the vortex magnetic field at the SC film surface
\cite{degennes}-\cite{abrik} is
\begin{eqnarray}
B_z^s (\rho)=\frac{\phi_0}{2\pi\lambda^2}\left[\frac{\lambda}{2\rho}
-\frac{\pi}{8} \left(H_0 (\frac{\rho}{2\lambda})
-N_0(\frac{\rho}{2\lambda})\right)\right]
\end{eqnarray}
where $H_0(x)$ and $N_0(x)$ are the zero order Struve and Neumann
functions respectively \cite{grad}.

\section{The Pinning Potential and Energy of FCD-SV System}

The interaction between a superconducting vortex and a non
magnetic columnar defect was first considered by Mkrtchyan and
Schmidt \cite{MS} and later in the work of Buzdin {\it{et.al.}}
\cite{buzdin1}-\cite{buzdin5}. In these studies, it was shown that
the pinning potential $U_p$ created by a non magnetic columnar
defect of radius $R>\sqrt{2}\xi$ is

\[U_p (\rho)=\left\{\begin{array}{rll}
           & &-\epsilon_0 \ln(\frac{R}{\sqrt{2}\xi}), \hspace{0.9in} \rho<R\\
       & &\epsilon_0 \ln \left[1 -\left(\frac{\sqrt{2} R}{\sqrt{2}
      \rho +\xi}\right)^2 \right], \hspace{0.1in} R<\rho<\lambda
         \end{array}\right.
\]

\noindent where $\xi$ is the SC coherence length and $\epsilon_0
=\frac{\phi_0^2}{16\pi^2 \lambda}$ is the energy scale of the
vortex self interaction.

If the columnar defect is ferromagnetic, then an extra contribution
to the pinning would appear due to the interaction between the
superconductor and ferromagnet. In the presence of vortices in the
superconductor, the total energy of the system is made up of five
different contributions and can be written as

\begin{eqnarray}
U=U_{sv} +U_{vv} +U_p +U_{mv} +U_{mm}
\end{eqnarray}
\noindent
where $U_{sv}$ is the energy of $N$ non-interacting vortices,
$U_{vv}$ is the vortex-vortex interaction, $U_{mv}$ is the
interaction energy between the FM and the SC, and $U_{mm}$ is
the FM dot self interaction. In \cite{ours}, it was shown that
the total energy of the system can be written as:
\begin{eqnarray}
U=\int \left[\frac{n_s \hbar^2}{8m_e}\left(\nabla\chi\right)^2
-\frac{n_s \hbar e}{4m_ec}\left(\nabla\chi \cdot
{\bf{A}}\right) -\frac{1}{2} \bf{M}\cdot\bf{B}\right]d^3x
\label{eqn11}
\end{eqnarray}
\noindent
where $c$ is the speed of light. The vectorial quantities
$\bf{A}$, and $\bf{B}$ are the total vector potential and magnetic
field due to the vortices and the ferromagnetic columnar defect.

In Ref.\onlinecite{buzd}, Buzdin considered the problem of vortex
craetion in a superconductor penetrated by non-magnetic columnar
defects and placed in external magnetic field. He found that at high
magnetic field giant vortices are more energy favorable when the
columnar defect is rather thick. The case of rather thick defects
$R/\lambda \sim 1$ is experimentally more interesting and easier to
realize. Marmorkos {\it{et.al.}} \cite{peeters} showed the
possibility of formation of a giant vortex with multiple flux quanta
around the magnetic dot instead of singly quantized vortices.
Therefore, I will assume that $\xi \ll R \le \lambda$ which make the
result of Ref.\onlinecite{buzd} regarding the formation of giant
vortices applicable to our problem.

The phase gradient of the SC order parameter in the presence of a giant
superconducting vortex with vorticity $\nu$ is
${\bf{\nabla}}\chi=\nu \frac{({\bf{\rho}}
-{\bf{\rho}}_0)\times \hat{z}}{|{\bf{\rho}} -{\bf{\rho}}_0|^2}$,
where ${\bf{\rho}}_0$ is the location of the vortex. The total
energy for a system consisting of a superconducting vortex and a
ferromagnetic columnar defect is

\begin{eqnarray}
U(\rho_0) &=& \nu^2 \epsilon_0 \ln(\frac{\lambda}{\xi}) +\nu^2 U_p(\rho_0)
\label{E1v}\\
& &-2\nu \epsilon_m \frac{R}{\lambda} \int_0^\infty
\frac{J_0(q\rho_0)J_1(qR)(1- e^{-\frac{qL}{2}})}{q(1+2\lambda q)}dq
\nonumber
\end{eqnarray}
where $\epsilon_m =M \phi_0 \lambda$ and $U_{mm}$ is the
self interaction of the FCD is ignored since it does not
affect the superconducting state. It is clear that if the
magnetization of the columnar defect is large enough so that
the interaction term becomes larger than other terms in
(\ref{E1v}) then the vortex can appear spontaneously in
the superconductor. Erdin {\it{et.al.}} \cite{ours} showed that
spontaneous creation of vortices takes place most easily in the
proximity of the superconducting transition temperature $T_s$.

The interaction term in Eq.(\ref{E1v}) depends on $L$. In the limit of
$L/R \ll 1$, the interaction term in Eq.(\ref{E1v}) reduces to the
result obtained for infinitely thin magnetic dot.\cite{ours} In the
remainder of this work, I will assume $L\rightarrow \infty$.
Numerical minimization of the total energy $U$ sets $\rho_0=0$,
hence the vortex center must be on the axis of the FCD.
\begin{figure}[ht]
  \centering
  \includegraphics[angle=0,width=3.5in,totalheight=2.5in]{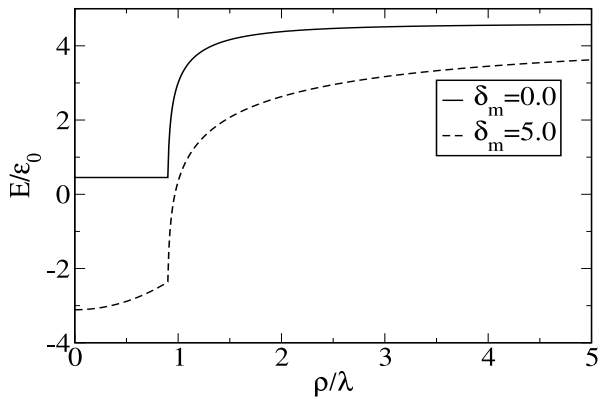}
  \caption{The total energy of the CD-SV system for
  $\lambda=1000$nm, $\xi=10$nm and the radius $R=900$nm.. The solid line
    is for the case when the CD is non-magnetic while the dashed one is
    for ferromagnetic CD.}
  \label{fig2}
\end{figure}

The energy of a system of a singly quantized vortex $\nu=1$ coupled
to a columnar defect is shown in Fig.~\ref{fig2}. The solid line
represents the energy of the system when the columnar defect is
non-magnetic, while the dashed line shows the energy of the system
when the defect is ferromagnetic. Note that the pinning force,
$-{\bf{\nabla}} U(\rho)$, is not zero for $\rho <R$ in contrast with
the case of non-magnetic columnar defect.

\begin{figure}[ht]
  \centering
  \includegraphics[angle=0,width=3.5in,totalheight=2.5in]{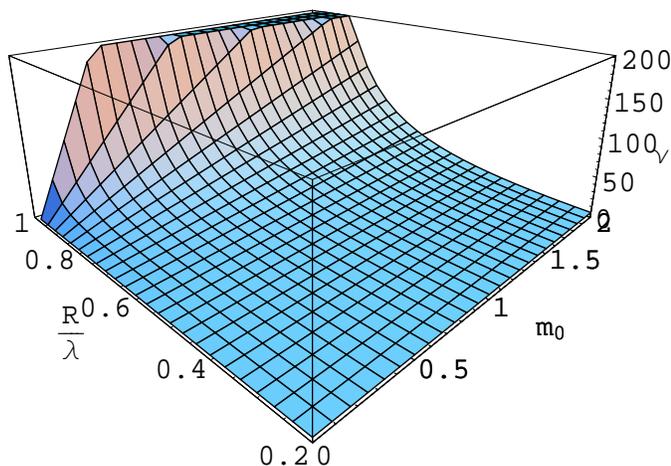}
  \caption{(Color Online) The plot of $\nu$ as a function of $m_0$
    and $R/\lambda$.}
  \label{fig4}
\end{figure}

\begin{figure}[ht]
  \centering
  \includegraphics[angle=0,width=3.5in,totalheight=2.5in]{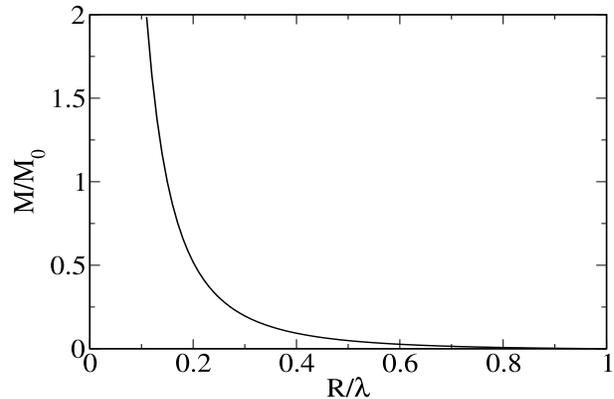}
  \caption{The curve represents the threshold value of the magnetization
  of the FCD in units of $M_0=\phi_0/\pi\lambda^2$.
  All points above the curve represent a phase with
  vortices while the region under the curve is vortexless.}
  \label{fig3}
\end{figure}

For each value of $M$ there is a corresponding critical value of
$\nu$ which I call it here $\nu_{c}$. The minimization of
$U(\rho_0=0)$ with respect to $\nu$ yields

\begin{eqnarray}
\nu=\frac{16\pi m_0}{\ln(\frac{\lambda}{R})} \frac{R}{\lambda}
\int_0^\infty \frac{J_1(qR)}{q(1+2\lambda q)}dq \label{nueq}
\end{eqnarray}
where $m_0=M/M_0$ with $M_0=\phi_0/\pi \lambda^2$. The critical
value of $\nu$ is the closest integer to the value of $\nu$ given by
Eq.(\ref{nueq}). The vorticity of the spontaneously created giant
vortex is plotted in Fig.~\ref{fig4} as a function of $m_0$ and
$R/\lambda$. Note that at when $R/\lambda \ll 1$ then very large
values of $m_0$ are required for the spontaneous creation of a
vortex.
Note that for any fixed value of $\nu_c$, Eq.(\ref{nueq}) gives us
the critical value of the magnetization $M_c (\nu_c)$ at which a
vortex with vorticity $\nu_c$ appear as a function of the radius
$R$. The dependance of $M_c(\nu_c)$ on $R$ for $\nu_{c}=1$ is
depicted in Fig.~\ref{fig3}. The region under the curve in
Fig.~\ref{fig3} represents a vortexless phase while the one above
the curve represents phases with vortices.

Note that when the radius of the columnar defect is very close to
$\lambda$ giant vortices with large flux quanta are expected even at
not so large values of $m_0$. This is because when $R \approx
\lambda$, the vortex self energy goes to zero while its energy of
interaction with the columnar defect is not zero and negative;
therefore, giant vortices with large value of $\nu$ are energy
favorable.

The curve in Fig.~\ref{fig3} is the curve for the first phase
transition (No vortex Phase $\rightarrow$ One vortex phase), it
should be emphasized here that there is a discrete set of curves
above this curve that refer to a series of phase transitions between
phases with different number of flux quanta. The transition from a
vortexless phase to a phase with a single vortex for a system in
which $R=50$ nm and $\lambda=100$ nm occurs at the approximate value
of $M_c=66.85$ G.

\section{An SC film Pierced by a square array of FCDs}
Now, let us consider a superconducting thin film pierced by a square
array of $N\times N$ ferromagnetic columnar defects. Let each FCD be
infinitely long ($L\rightarrow\infty$) with a radius $\xi \ll R \le
\lambda$ and magnetization $M \ge M_c (\nu_c)$. The magnetization of
the array can be written in Fourier space as follows

\begin{eqnarray}
{\bf{M_K}}=4\pi^2 M R\frac{J_1 (qR)}{q} \delta (k_z) \hat{z} \sum_j
e^{\imath {\bf{q.\rho_j}}}
\label{magarray}
\end{eqnarray}
\noindent
where ${\bf{\rho}}_i$ is the location of the $i$-th FM
defect and the sum runs over all the lattice sites. The location of
the FCD is defined by two integers $l$ and $s$ as ${\bf{\rho}}_i=l a
\hat{x}+ s a\hat{y}$ where $a$ is the lattice spacing of the FCD
array. Solving Eq.(\ref{GLM}) for the magnetization (\ref{magarray})
yields

\begin{eqnarray}
\tilde{{\bf{A}}}_K^m &=& \frac{16\pi^2 \imath M R J_1 (qR)}{q(k_z^2 +q^2)}
\left(\pi \delta(k_z) -\frac{1}{q(1+2\lambda q)}\right)
\sum_j e^{\imath {\bf{q.\rho_j}}}\nonumber\\
& & \times ({\bf{q}}\times \hat{z})
\label{FRVParray}
\end{eqnarray}

Let {\bf{Q}} be a vector of the reciprocal lattice defined as
${\bf{Q}}=n {\bf{Q}}_1 +m {\bf{Q}}_2$ where $n$ and $m$ are two
integers with ${\bf{Q_1}}=\frac{2\pi}{a}\hat{x}$ and ${\bf{Q}}_2
=\frac{2\pi}{a}\hat{y}$. The $z$ component of the array's magnetic
field at $z=0$ is
\begin{eqnarray}
B_z (x,y, z=0)=16\pi^2 n_0 M \lambda R \sum_{\bf{Q}}\frac{J_1 (Q
R)}{1+2\lambda Q} e^{-\imath {\bf{Q.\rho}}} \label{arrayfield}
\end{eqnarray}
\noindent
where $n_0=1/a^2$ is the density of columnar defects in the array.
The summation over the defect positions is calculated with the help
of the identity
\begin{eqnarray}
\sum_i \int f(q) e^{\imath {\bf{q.}}{\bf{\rho}}_i}
\frac{d^2q}{(2\pi)^2}= \sum_{\bf{Q}} n_0 f({\bf{Q}}) \label{kop1}
\end{eqnarray}

The magnetic flux which penetrate the superconductor must remain
constant; therefore, the number of vortices and the number of
antivortices that are spontaneously created in the superconductor
must be equal. In the absence of external magnetic field, the
equilibrium configuration is the one with vortices pinned at the
axes of the FCDs and antivortices are located at the centers of the
unit cells of the FCD lattice. The square lattice of antivortices is
identical to that of the vortices with a shift by
${\bf{a}}/2=(a/2,a/2)$. Note that in the case of a single columnar
defect, the antivortex is repelled (pushed far away) from the FCD;
therefore, it was not taken into account in the calculations.
However, in the case when the SC is penetrated by an array of FCD in
the absence of external magnetic field, one must take antivortices
into account on equal footing with vortices. In equilibrium, the
force acting on any flux line (FL) in the superconductor is zero.
Pinning forces appear as soon as this equilibrium is disturbed e.g.
by passing an electric current to the superconducting film. The
remaining part of this section will focus on the problem of
calculating these forces. The vortex and antivortex lattices are
both regular and periodic. Therefore, if a small current is applied
to the superconductor then all vortices will move together in one
direction and all antivortices will move together in the opposite
direction. Let us assume that the vortex lattice is displaced with
respect to its equilibrium position by a small amount
$\Delta{\bf{\rho}}=(x,y)$ then the lattice of antivortices will be
displaced by an amount $-\Delta{\bf{\rho}}$. Considering the case
when $M$ is just above $M_c(\nu_c=1)$. The flux line self energy and
the FL-FL interaction energy together can be written using
Eq.(\ref{eqn11}) as
\begin{widetext}
\begin{eqnarray}
E_{vv} =\frac{\mathcal{A} n_0^2 \phi_0^2}{2\pi}\sum_{\bf{Q}}
\frac{1}{Q(1+2\lambda
Q)}\left[1-\cos({\bf{Q}}\cdot(\frac{{\bf{a}}}{2}-2\Delta
{\bf{\rho}}))\right] \label{HVA}
\end{eqnarray}
where $\mathcal{A}$ is the area of the SC film. The energy of
interaction between the flux lines and the ferromagnetic array is
\begin{eqnarray}
E_{mv}=-2\pi \mathcal{A} n_0^2 M R\lambda \sum_{\bf{Q}}
\frac{J_1(QR)}{Q^2 (1+2\lambda Q)}\left[\cos({\bf{Q}}\cdot \Delta
{\bf{\rho}}) -\cos({\bf{Q}}\cdot (\frac{{\bf{a}}}{2}-\Delta
{\bf{\rho}})\right] \label{HMA}
\end{eqnarray}
\end{widetext}
To simplify the calculations, I use
\begin{eqnarray}
\sum_{\bf{Q}} \Rightarrow \mathcal{A}_0 \int \frac{d^2Q}{(2\pi)^2}
\end{eqnarray}
where $\mathcal{A}_0=a^2$ is the area of the array's unit cell.
Therefore, I find the energy per flux line is now dependent on the
displacement $\Delta{\bf{\rho}}$ as follows
\begin{eqnarray}
\mathcal{E}_{o} &=& \frac{\phi_0^2}{8\pi^2} \int_0^\infty
\frac{\left[1-J_0(q|\frac{{\bf{a}}}{2}
-2\Delta{\bf{\rho}}|)\right]}{1+2\lambda q} dq \\
& &- M\phi_0 R\int_0^\infty
\frac{J_1(qR)\left[J_0(q|\Delta{\bf{\rho}}|)
-J_0(q|\frac{{\bf{a}}}{2} -\Delta{\bf{\rho}}|)\right]}{q(1+2\lambda
q)}dq \nonumber
\label{new2}
\end{eqnarray}
Note that the number of FCD is $N^2$ while the number of flux lines
is $2N^2$. The pinning force acting on any vortex or antivortex can
now be calculated using ${\bf{f}}_p=-{\bf{\nabla}}
U(\Delta{\bf{\rho}})$ where $U(\Delta{\bf{\rho}})=\mathcal{E}_o
-\mathcal{E}_e$ with $\mathcal{E}_e$ is a constant equals to the
energy of the system in equilibrium. Note that I did not include the
term due to the Mkrtchyan and Schmidt in the result I obtained in
Eq.(\ref{new2}) because it is rather straightforward to calculate
its effect.

\section{Conclusion}
In conclusion, I studied the interaction between superconducting
vortices in a thin film pierced by a ferromagnetic columnar defect.
I calculated the magnetic field of the FCD in the presence of the SC
and the distribution of screening currents in the superconductor. If
the magnetization of the FCD exceeds a critical value, then the FCD
interaction with the vortex will overcome the vortex self energy
leading to the spontaneous creation of vortices in the
superconductor. I showed that vortex pinning is strongly enhanced
due to the contribution from the interaction between the magnetic
field of vortices and the magnetization of the FCD outside the plane
of the SC and its close proximity. This extra contribution to vortex
pinning is a major difference between this study and other studies
in the literature which focused on dipole-like or two-dimensional
ferromagnetic structures. These results were generalized to include
the case when the SC film is pierced by an array of FCD. A detailed
analysis of the dynamical properties of this system will be reported
elsewhere.

I would like to thank V. L. Pokrovsky, W. M. Saslow and D. G. Naugle
for useful discussions. I acknowledge partial support by the NSF
grant DMR 0321572 and DOE grant DE-FG03-96ER45598 during my stay at
Texas A\&M University. I also would like to thank P. R. Montague at
Baylor College of Medicine for his support and encouragement.

\end{document}